\documentclass[reprint,twocolumn,superscriptaddress,preprintnumbers,amsmath,amssymb,aps,prd,floatfix]{revtex4-2}

\usepackage[utf8]{inputenc}
\usepackage{enumerate}

\usepackage{mathtools}
\usepackage{amsfonts}
\usepackage{mathrsfs}
\usepackage{bm}
\usepackage[normalem]{ulem}
\usepackage{bbold}
\usepackage{braket}
\usepackage{slashed}
\usepackage{dsfont}
\usepackage{float}
\usepackage{dcolumn}
\usepackage{amssymb,amsmath}
\usepackage{titlesec}

\usepackage{graphicx}
\usepackage{color}
\usepackage{array}

\usepackage{placeins}
\usepackage{booktabs}

\usepackage{xspace}
\usepackage{hyperref}
\usepackage[nameinlink]{cleveref}
\usepackage{bookmark}
\usepackage{units}

\def\Eq#1{Eq.~\labelcref{#1}}

\def\eq#1{\labelcref{#1}}
\def\Fig#1{Fig.~\labelcref{#1}}

\setkeys{Gin}{width=0.48\textwidth}

\usepackage{booktabs}
\usepackage{multirow}

\newcolumntype{C}{>{$}c<{$}}
\AtBeginDocument{
	\heavyrulewidth=.08em
	\lightrulewidth=.05em
	\cmidrulewidth=.03em
	\belowrulesep=.65ex
	\belowbottomsep=0pt
	\aboverulesep=.4ex
	\abovetopsep=0pt
	\cmidrulesep=\doublerulesep
	\cmidrulekern=.5em
	\defaultaddspace=.5em
}

\graphicspath{{./figures/}}

\usepackage{xifthen}
\usepackage{xcolor}

\newcommand{\gettitle}{Real-time evolution of critical modes in the QCD phase diagram}

\newcommand{\getDalianAffiliation}{\affiliation{School of Physics, Dalian University of Technology, Dalian, 116024, P.R. China}}

\newcommand{\getGiessenAffiliation}{\affiliation{Institut f\"ur Theoretische Physik, Justus-Liebig-Universit\"at Gie\ss en, 35392 Gie\ss en, Germany}}

\newcommand{\getHeidelbergAffiliation}{\affiliation{Institut f\"ur Theoretische Physik, Universit\"at Heidelberg, Philosophenweg 16, 69120 Heidelberg, Germany}}

\hypersetup{
	colorlinks,
	linkcolor={blue!75!black},
	citecolor={blue!75!black},
	urlcolor={blue!75!black}, 
	pdftitle={\gettitle},
	pdfauthor={},
	pdfkeywords={}
	{} 
	bookmarksopen=true,
	bookmarksopenlevel=2,
	bookmarksnumbered=true
}

\begin{document}
\title{\gettitle}

\author{Yang-yang Tan}
\getDalianAffiliation

\author{Shi Yin}
\getGiessenAffiliation

\author{Yong-rui Chen}
\getGiessenAffiliation

\author{Chuang Huang}
\getHeidelbergAffiliation

\author{Wei-jie Fu}
\email{wjfu@dlut.edu.cn}
\getDalianAffiliation

\begin{abstract}

A QCD-assisted relaxation dynamic model for the critical mode of the critical end point (CEP) in the QCD phase diagram is developed, which allows us to investigate the critical slowing down effect quantitatively in the QCD phase diagram, especially in the proximity of the CEP, without any phenomenological parameters. The relaxation time from nonequilibrium to equilibrium in the QCD phase diagram is extracted from the Langevin simulations of the QCD-assisted relaxation dynamic model. It is found that in a narrow region along the phase boundary radiated from the CEP, the relaxation time is enhanced significantly. Outside this narrow region, the relaxation time drops drastically, which implies that the dynamic critical region is small in the QCD phase diagram. We also find that the effects of critical slowing down are mild on the chemical freeze-out curves.

\end{abstract}

\maketitle

\emph{Introduction.--}
Studying the QCD phase diagram plays a pivotal role in exploring properties of strongly interacting matter under extreme conditions \cite{Stephanov:2007fk, STAR:2010vob, Luo:2017faz, Andronic:2017pug, Bzdak:2019pkr, Chen:2024aom, Dupuis:2020fhh, Fu:2022gou}. The QCD phase diagram is characterized by the critical end point (CEP) of possible existence \cite{Stephanov:1998dy, Stephanov:1999zu}, that is the end point of the first-order phase transition line in the regime of high baryon chemical potential, or equivalently high baryon densities. Recent years have seen significant progress in the studies of QCD phase diagram, e.g., the estimate of location of the CEP in the phase diagram from direct computations of functional QCD \cite{Fu:2019hdw, Gao:2020fbl, Gunkel:2021oya} or extrapolations of lattice QCD based on Yang-Lee edge singularities \cite{Clarke:2024ugt, Borsanyi:2025dyp}. Fluctuation observables, such as the high-order cumulants of net-proton (baryon) number distributions, are employed to search for the CEP in heavy-ion collision experiments \cite{STAR:2020tga, STAR:2021fge, STAR:2022vlo, STAR:2022etb, STAR:2025zdq, Stephanov:2011pb, Fu:2015naa, Fu:2015amv, Fu:2016tey, Fu:2021oaw, Fu:2023lcm, Lu:2025cls}, where a peak structure of the kurtosis of baryon number fluctuations as a function of the collision energy is expected, as the fireball in heavy-ion collisions evolves near the CEP \cite{Stephanov:2011pb, Fu:2023lcm, Lu:2025cls}.

Another intricacy arises from the effect called the critical slowing down \cite{Berdnikov:1999ph}. As the system is evolving in the proximity of CEP, the correlation length $\xi$ increases significantly due to the nature of second-order phase transition for the CEP. Since the curvature of effective potential of the critical mode, i.e., the sigma mode, behaves as $\sim 1/\xi^2$, the relaxation time from nonequilibrium to equilibrium would be enhanced remarkably near the CEP. Consequently, the nonequilibrium effects would come into play, for instance drastically modifying the high-order fluctuations due to the memory effects \cite{Mukherjee:2015swa}.

To resolve the problems mentioned above, one in principle needs to simulate the real-time dynamics from microscopic interactions, e.g., QCD for a many-body system, that is, however, a formidable task apparently for the moment. It turns out that one can simplify the situations by starting from, instead of microscopic, but rather mesoscopic dynamics. For instance, the dynamic models summarized in \cite{Hohenberg:1977ym} have provided us with valuable information about the real-time dynamics of critical phenomena in phase transitions. It is thought that the $O(4)$ chiral phase transition of QCD is classified as Model G of dynamic universality classes \cite{Rajagopal:1992qz}, and the CEP as Model H \cite{Son:2004iv}.

Although the mesoscopic dynamic models are powerful in studying dynamic universal properties of critical phenomena inside the critical region, e.g., the dynamic critical exponent \cite{Hohenberg:1977ym}. This is, however, not the case outside the critical region when non-universal properties are concerned, since dynamic models usually rely on phenomenological parameters, such as various transport coefficients, which cannot be computed in the models themselves, but rather have to be calculated from microscopic interactions. Therefore, it is a long dream in the studies of real-time dynamics to combine the microscopic with mesoscopic, or even the macroscopic approaches, e.g., the hydrodynamics. 

Renormalization group (RG) is a tailor-made method to resolve the physics of different scales. In particular, the functional renormalization group (fRG) has been successfully used in the studies of both QCD at finite temperature and densities \cite{Fu:2019hdw, Braun:2020ada, Braun:2023qak, Fu:2024rto, Dupuis:2020fhh, Fu:2022gou} and dynamic models \cite{Canet:2006xu, Canet:2011wf, Mesterhazy:2013naa, Bluhm:2018qkf, Chen:2023tqc, Roth:2023wbp, Tan:2024fuq, Chen:2024lzz, Roth:2024rbi, Roth:2024hcu}. In this letter, we would like to combine the microscopic QCD and the mesoscopic dynamic models within the fRG approach, which allows us to develop a QCD-assisted relaxation dynamics for the critical mode of CEP. In the QCD-assisted relaxation dynamics, the phenomenological parts are completely computed from QCD at finite temperature and densities. Therefore, one is able to investigate the properties of critical slowing down quantitatively in the QCD phase diagram, specifically near the CEP.

In this letter, we first describe the relaxation dynamics of the critical mode with the Schwinger-Keldysh effective action. The QCD-assisted Langevin transport model is then discussed in detail. After that we simulate the QCD-assisted Langevin transport in 3+1 dimensions, from which the relaxation time of the critical mode in the QCD phase diagram can be extracted. More details about the QCD-assisted Langevin transport model and its simulations are presented in the supplement. Our results offer a novel perspective on the critical slowing down near the CEP in the QCD phase diagram.

\emph{Relaxation dynamics of the critical mode.--}
The non-equilibrium relaxation evolution of a critical mode, e.g., the $\sigma$ mode in QCD phase transitions, can be well described by Model A, that is a purely dissipative relaxation model classified in the seminal review about dynamical models in \cite{Hohenberg:1977ym}. The related Schwinger-Keldysh (SK) effective action reads
\begin{align}
  \Gamma[\phi_c,\phi_q]=&\int \mathrm{d}^4 x\Big(Z_\phi^{(t)}\sigma_q\partial_t\sigma_c-Z^{(i)}_{\phi}\sigma_q\partial_i^2\sigma_c\nonumber\\[2ex]
  &+2U'(\sigma_c)\sigma_{q}-2Z_\phi^{(t)}T\sigma_q^2\Big),\label{eq:action_sigma_Keldysh_main}
\end{align}
with the field $\phi=\{\sigma\}$, $\partial_t \equiv \partial/ \partial t$ and $\partial_i \equiv \partial/ \partial x^i$ ($i=1,2, \cdots d$). Here the spatial dimension $d=3$ is chosen and a summation for $i$ in $\partial_i^2$ is assumed. The subscripts `$c$' and `$q$'  stand for the `classical' and `quantum' fields, respectively, see, e.g. \cite{Tan:2021zid, Chen:2023tqc, Tan:2024fuq} for more details. Discussions about the functional renormalization group with the Schwinger-Keldysh path integral can be found in, e.g., \cite{Berges:2008sr, Pawlowski:2015mia, Tan:2021zid}. In effective action \labelcref{eq:action_sigma_Keldysh_main},  $Z_\phi^{(t)}$ and $Z^{(i)}_{\phi}$ are the temporal and spatial wave functions, respectively. $U(\sigma_{c})$ denotes the effective potential and $U'(\sigma_{c})$ the derivative, that arises from the difference between the effective action on the two different time branches on the SK contour. The last term quadratic in $\sigma_q$ represents a Gaussian white noise with temperature $T$. Note that the temporal wave function $Z_\phi^{(t)}$ appears as the coefficient of noise term due to the fluctuation-dissipation theorem, see \cite{Tan:2024fuq} for more details.

The effective action in \Eq{eq:action_sigma_Keldysh_main} is expanded up to the quadratic term in $\phi_q$ field, which is also well known as the semi-classical approximation. The semi-classical effective action is equivalent to a description of Langevin transport. The counterpart of \Eq{eq:action_sigma_Keldysh_main} in terms of Langevin equation reads
\begin{subequations} 
\label{eq:Langevin_overdamped2}
\begin{align}
  Z_\phi^{(t)}\partial_t\sigma-Z^{(i)}_{\phi}\partial_i^2\sigma+U'(\sigma)=\xi\,,\label{eq:Langevin-Eq}
\end{align}
with the correlation of the Gaussian white noise 
\begin{align}
  \langle \xi(t,\bm x)\xi(t',\bm x')\rangle=2Z_\phi^{(t)}T\delta(t-t')\delta(\bm x-\bm x')\,.\label{eq:Langevin-correlation}
\end{align}
\end{subequations}
In \Eq{eq:Langevin_overdamped2} we have used $\sigma=\sigma_c/\sqrt{2}$ for simplicity.

\emph{QCD-assisted Langevin transport model.--}
In order to simulate the Langevin transport equation \labelcref{eq:Langevin_overdamped2}, one needs the inputs of three components: The potential $U(\sigma)$ or $U'(\sigma)$, the spatial wave function of the $\sigma$ mode $Z^{(i)}_{\phi}$ and the temporal wave function $Z_\phi^{(t)}$, respectively. Note that the first two quantities describe static properties of the system, thus can be calculated directly from, e.g., Euclidean field theory. In contradistinction, the temporal wave function is essentially a dissipation coefficient and depicts the dynamic properties, so it can only be computed from real-time correlation functions in Minkowski field theories, e.g., the SK field theory, or equivalently from the real-time correlation functions continued analytically from the Euclidean ones.

In this work, we adopt the three ingredients necessary in \Eq{eq:Langevin_overdamped2}, calculated from QCD at finite temperature and densities within the fRG approach \cite{Fu:2019hdw, Fu:2024rto}. This provides us with a unified method to obtain both the static and dynamic properties of hot QCD matter. In the fRG approach to QCD, quantum fluctuations of quarks, gluons and ghosts, where usually the Landau gauge is chosen in functional QCD, are integrated in successively with the evolution of the renormalization group (RG) scale from the ultraviolet (UV) to infrared (IR). Moreover, RG also provides us with a tailor-made approach to deal with transition of degrees of freedom (d.o.f.) in the low energy QCD. When RG scale is lowered down such that the system enters into the regime of nonperturbative QCD, e.g., $\lesssim 1$ GeV, a finite gluon mass develops, which entails the glue sector, including the gluons and ghosts, decouples from the system \cite{Alkofer:2000wg, Fischer:2008uz}. Even more, as the chiral symmetry is broken spontaneously, collective composite d.o.f., e.g., the pion mesons and the sigma mode $\phi=\{\sigma, \pi\}$, take over the fundamental d.o.f of $\varphi=\{A, c, \bar{c}, q, \bar{q}\}$ and behave as the most active d.o.f. in the low energy QCD. Here, $A, c, \bar{c}, q, \bar{q}$ stand for the gluon, ghost, anti-ghost, quark and anti-quark, respectively. This transition of d.o.f. is well described in the fRG approach by virtue of a mature technique called dynamical hadronization \cite{Gies:2001nw, Gies:2002hq, Pawlowski:2005xe, Fu:2019hdw}. It is found that there is no double counting with the dynamical hadronization, that has been investigated in detail in \cite{Braun:2014ata}. 

%
\begin{figure*}[t]
  \includegraphics[width=0.49\textwidth]{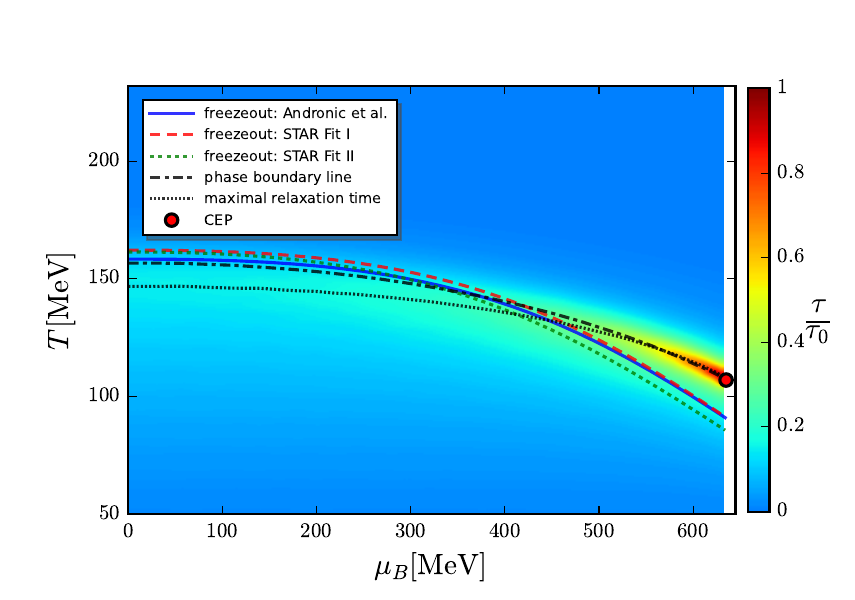}
  \hfill
  \includegraphics[width=0.49\textwidth]{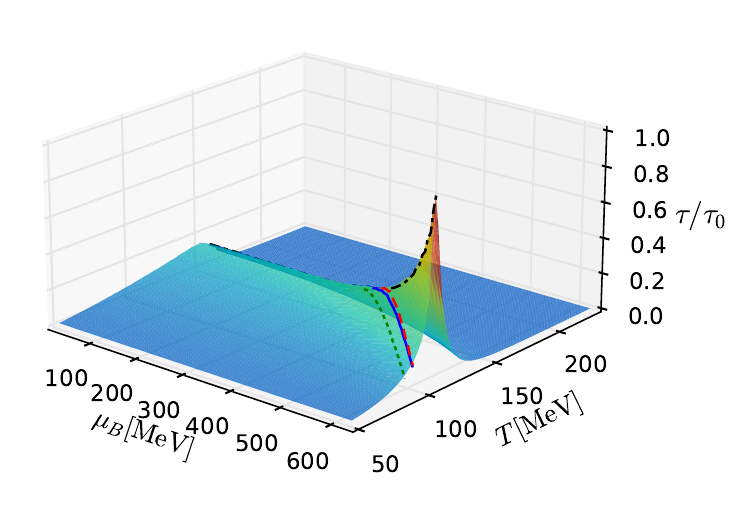}
\caption{\emph{Left panel:} Heatmap of the reduced relaxation time $\tau/\tau_0$ in the QCD phase diagram. The relaxation time is normalized by the maximal relaxation time $\tau_0$ in this plane, which is located at a point near the CEP, denoted by the red dot. Three typical chemical freeze-out curves used also in \cite{Fu:2021oaw} are shown. The black dot-dashed line denotes the phase boundary of crossover obtained in QCD within the fRG \cite{Fu:2019hdw}. The black dotted line stands for the line of the maximal relaxation time along the temperature direction at a fixed baryon chemical potential. \emph{Right panel:} 3D plot of the left panel.}\label{fig:relaxtime}
\end{figure*}
%

Consequently, one is able to compute correlation functions for not only the fundamental fields $\varphi$ but also the composite fields, e.g., $\phi$, in the fRG approach to QCD. All these correlation functions can be derived from a unified effective action $\Gamma [\Phi]$ with $\Phi=\{\varphi, \phi\}$. For instance, the effective potential in \Eq{eq:action_sigma_Keldysh_main} is obtained as follows
\begin{align}
  U'(\sigma)=\frac{\delta \Gamma[\Phi]}{\delta \sigma }\bigg|_{\substack{\sigma(x)=\sigma\\ \tilde\Phi=\tilde\Phi_{\mathrm{EoM}}}}\,,\label{eq:UsigmaGam}
\end{align}
where we consider a coordinate-independent potential and $\tilde\Phi$ denotes all the other fields except $\sigma$, which are on their respective equations of motion (EoM). In the same way, the wave functions can be obtained from the two-point correlation function of mesons, i.e.,
\begin{align}
  \Gamma^{(2)}_{\sigma \sigma}=\frac{\delta^2 \Gamma[\Phi]}{\delta \sigma \delta \sigma}\bigg|_{\Phi=\Phi_{\mathrm{EoM}}}\,.\label{eq:Gam2sig2}
\end{align}
Then, one finds for the spatial wave function 
\begin{align}
  Z^{(i)}_{\phi}=\frac{\partial \Gamma^{(2)}_{\sigma \sigma}(p_0, \bm{p})}{\partial \bm{p}^2}\bigg|_{\substack{p_0=0\\ \bm{p}=0}}\,.\label{eq:Zphii}
\end{align}
As discussed above, the computation of the temporal wave function is more involved. The Euclidean correlation function in \Eq{eq:Gam2sig2} has to be continued analytically to the retarded correlation function in the Minkowski spacetime, that is,
\begin{align}
  \Gamma^{(2)}_{\sigma \sigma,\mathrm{R}}(\omega, \bm{p})=\lim_{\epsilon\to 0^+}\Gamma^{(2)}_{\sigma \sigma}\big(p_0=-\mathrm{i}(\omega+\mathrm{i}\epsilon),\bm{p}\big)\,.\label{eq:AnalyContiGamsig2}
\end{align}
Note that in the fRG approach to QCD, the analytic continuation is done on the level of the flow equations, which allows us to perform numerical computation in the Minkowski spacetime directly, and thus does not suffer from subtlety of inverse problems if the continuation is done on, e.g., numerical data. Details about the calculations of real-time correlations within the fRG approach to QCD can be found in Ref.~\cite{Fu:2024rto}. Consequently, the temporal wave function can be obtained from the imaginary part of the retarded two-point correlation function of sigma mode, i.e.,
\begin{align}
  Z_\phi^{(t)}=\lim_{\omega\to 0}\frac{\partial}{\partial \omega}(-)\mathrm{Im}\,\Gamma^{(2)}_{\sigma \sigma,\mathrm{R}}(\omega, |\bm{p}|=p_L)\,,\label{eq:Zphit}
\end{align}
where the long time limit $\omega\to 0$ is implemented. In \Eq{eq:Zphit} the spatial momentum is chosen to be a value $|\bm{p}|=p_L$, corresponding to the size of the system. A typical value $p_L=10$ MeV corresponds to the length scale $1/p_L \sim 20$ fm, that is comparable to the typical size of fireball in heavy-ion collisions. We also change a bit the value of $p_L$, and we are interested in the relative change of the relaxation time in the QCD phase diagram, which is found to be insensitive to the specific value of $p_L$. Note that one is not allowed to adopt the limit $|\bm{p}|=p_L \to 0$, since it is found that for the sigma mode in \Eq{eq:Zphit}, one has $\mathrm{Im}\,\Gamma^{(2)}_{\sigma \sigma,\mathrm{R}} \sim 1/ |\bm{p}|$ in the limit of $|\bm{p}| \to 0$, which is discussed in detail in the supplement. It should be noted that a finite value of the imaginary part of $\Gamma^{(2)}_{\sigma \sigma,\mathrm{R}}$ arises from the Landau damping effect, i.e., the particle-hole process of quarks near the Fermi surface, which has been discussed in detail in \cite{Fu:2024rto}. The critical mode exchanges energy of slow modes with the heat bath through the Landau damping.

\emph{Simulations of QCD-assisted Langevin transport in 3+1 dimensions.--}
With the quantities in \labelcref{eq:UsigmaGam,eq:Zphii,eq:Zphit} computed within the fRG approach to QCD at finite temperature $T$ and baryon chemical potential $\mu_B$ \cite{Fu:2019hdw}, one is able to simulate the Langevin equation \labelcref{eq:Langevin_overdamped2} in 3+1 dimensions, see also related classical-statistical simulations \cite{Aarts:2001yx, Batini:2023nan, Schweitzer:2021iqk, Chattopadhyay:2024jlh}. We choose a three-dimensional spatial lattice of periodic boundary conditions with lattice constant $a$ and size $N_L^3$. Dependence of simulation results on the lattice spacing $a$ and extent $N_L$ are investigated in detail and the relevant results are presented in the supplement. It is found convergent results are obtained with $a=0.1$ fm and $N_L=32$, where the continuum limit is accessible and the effects of finite volume are negligible. The Langevin equation \labelcref{eq:Langevin_overdamped2} is implemented as a stochastic partial differential equation in Julia on GPUs \cite{Julia-2017,besard2018juliagpu}, where the time step is chosen adaptively with the control of relative error. It is found that the relative error $\sim 10^{-2}$ is sufficient to resolve the temporal evolution. The Gaussian white noises are generated by $N_L^3$ independent generators of random numbers with normal distribution.

In the simulations it is more convenient to work with the renormalized quantities, e.g., the renormalized sigma field $\bar\sigma=\sqrt{Z^{(i)}_{\phi}}\sigma$, cf. more details in the supplement. The average of the sigma field on the lattice for one sample in the ensemble reads
\begin{align}
  \tilde\sigma(t)=\frac{1}{N_L^3}\sum_{\bm x} \bar \sigma(t,\bm x)\,.\label{}
\end{align}
Then, the ensemble average $\langle\tilde\sigma(t)\rangle$ is obtained by averaging $\tilde\sigma(t)$ over $N_{\mathrm{ens}}$ samples in the ensemble. The error arising from a finite sample number $N_{\mathrm{ens}}$ is investigated in the supplement, and it is found that the errors are very small when the sample number is increased up to, e.g., $N_{\mathrm{ens}}=256$. Different schemes for the initial conditions of sigma field are adopted for the studies of critical mode relaxation from nonequilibrium to equilibrium, e.g., spatially homogeneous initial field configurations, field configurations of the Gaussian distribution, quenching from another equilibrium state, etc. It is found that the relaxation time is independent of the choice of initial conditions, see the supplement for more detailed discussions. The relaxation time $\tau$ can be extracted by fitting the ensemble-averaged sigma field with an exponential decay, as follows 
\begin{align}
  \langle\tilde\sigma(t)\rangle=\mathcal{C} \mathrm{e}^{-\frac{t}{\tau}}+\langle\tilde\sigma\rangle_{\mathrm{eq}}\,,\label{eq:relaxation_fit}
\end{align}
where $\langle\tilde\sigma\rangle_{\mathrm{eq}}$ is the value in equilibrium and $\mathcal{C}$ is a constant.

%
\begin{figure}[t]
\hspace{-1.cm}
\includegraphics[width=0.45\textwidth]{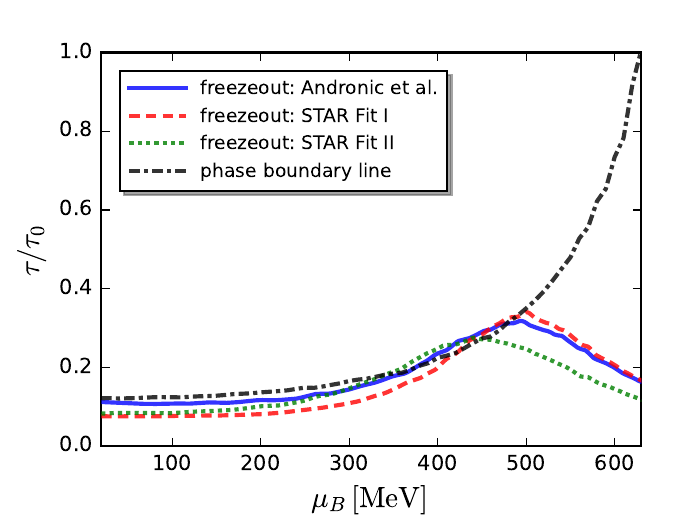}
\caption{Relaxation time as a function of the baryon chemical potential along the phase boundary and the three chemical freeze-out curves, respectively, as shown in \Fig{fig:relaxtime}.}
\label{fig:freezeout}
\end{figure}
%

\emph{Relaxation time of the critical mode in the QCD phase diagram.--}
In the left panel of \Fig{fig:relaxtime} we show the reduced relaxation time $\tau/\tau_0$ in the QCD phase diagram. The relaxation time is normalized by some reference relaxation time $\tau_0$, which here is chosen to be the maximal value in the part of phase diagram shown in \Fig{fig:relaxtime}. Apparently, this is a point near the CEP. The red dot stands for the location of CEP as shown in \Eq{eq:CEP}. In the phase diagram we also show the phase boundary obtained in \cite{Fu:2019hdw}, which is depicted through the pseudo-critical temperature of the renormalized light quark condensate as a function of the baryon chemical potential, see \cite{Fu:2019hdw} for more details. Moreover, we also show three typical freeze-out curves, that are also used in \cite{Fu:2021oaw, Fu:2023lcm}. The freeze-out: Andronic et al. is obtained from the freeze-out data in \cite{Andronic:2017pug}. The freeze-out: STAR Fit I and freeze-out: STAR Fit II are obtained from the STAR freeze-out data \cite{STAR:2017sal}, while for the latter some flawed data from general considerations are dropped, see \cite{Fu:2021oaw} for more details. The 3D plot of the relaxation time in the phase diagram is also shown in the right panel of \Fig{fig:relaxtime}.

The location of CEP is found to be 
\begin{align}
(T_\mathrm{CEP},\mu_{B_{\mathrm{CEP}}})=(107,635)\,\mathrm{MeV}\,, \label{eq:CEP}
\end{align}
in \cite{Fu:2019hdw}. Simulations of Langevin transport become difficult as we approach towards the CEP. Therefore, the computations are constrained in the regime of up to the baryon chemical potential $\mu_B=630$ MeV in this work. In this regime, the maximal relaxation time $\tau_0$ is found to take place at $(T,\mu_{B})\sim (108, 630)$ MeV, which is also used to normalize the relaxation time and leaves us with the reduced relaxation time $\tau/\tau_0$. In the left panel of \Fig{fig:relaxtime} we also show the line of maximal relaxation time, which is determined by the temperature related to the maximal relaxation time for each value of $\mu_{B}$. One can see that the line of maximal relaxation time is below the line of the phase boundary in the regime of small $\mu_{B}$, which is consistent with the nature of continuous crossover in this regime. The phase boundary and the line of maximal relaxation approach each other with the increasing $\mu_{B}$, and converge near the CEP.

Another interesting finding is that as the baryon chemical potential is increased up to, e.g., $\mu_{B}\gtrsim 400$ MeV, the freeze-out curves bend downwards and deviate from the phase boundary line. Consequently, the relaxation time decreases with $\mu_{B}$ rather than increases along the freeze-out curves in the region of large $\mu_{B}$, which is shown clearly in \Fig{fig:freezeout}. We also show the relaxation time along the phase boundary line in \Fig{fig:freezeout} for comparison, which increases with $\mu_{B}$ monotonically. 

The observation that the relaxation time on the chemical freeze-out curves is not increased significantly with the increase of $\mu_{B}$, is also attributed to a fact found from our calculations, that is, a small dynamical critical region. As shown in the left panel of \Fig{fig:relaxtime}, the red region in the proximity of CEP is very small. This is even clearer in the 3D plot in the right panel of \Fig{fig:relaxtime}, where one can see the relaxation time drops immediately once it deviates from the CEP. The small size of the dynamical critical region is consistent with the small size of the static critical region in QCD \cite{Braun:2023qak, Tan:2024fuq}. The fact that the chemical freeze-out curves are away from the dynamical critical region implies that the effects of critical slowing down are mild on the chemical freeze-out curves. Consequently, observables in heavy-ion collisions which are determined mainly at the chemical freeze-out, e.g., the baryon number fluctuations, do not suffer from the effect of critical slowing down severely, and thus can be described by hydrodynamic simulations combined with local equilibrium critical fluctuations \cite{Fu:2021oaw, Fu:2023lcm}.


\emph{Conclusions and outlook.--}
In this letter, we have developed a QCD-assisted relaxation dynamic model for the critical mode of CEP in the QCD phase diagram at finite temperature and baryon chemical potential. Both the static effective potential and the dynamic transport coefficient in the relaxation dynamic model are computed directly from the flow equations of Euclidean and Minkowski correlation functions in QCD within the fRG approach. This allows us to investigate the critical slowing down effect quantitatively in the QCD phase diagram, especially in the proximity of the CEP, without any phenomenological parameters.

The relaxation time from nonequilibrium to equilibrium in the QCD phase diagram is extracted from the Langevin simulations of the QCD-assisted relaxation dynamic model. It is found that in a narrow region along the phase boundary radiated from the CEP, the relaxation time is enhanced significantly. Outside this narrow region, the relaxation time drops drastically, which implies that the dynamic critical region is small in the QCD phase diagram. Moreover, we also compute the relaxation time on the chemical freeze-out curves. It is found that with the increase of the baryon chemical potential, the relaxation time is not increasing monotonically, but instead develops a peak structure and decreases in the regime of high baryon chemical potential. This is attributed to fact that the freeze-out curves bend down and deviate from the phase boundary in the regime of large baryon chemical potential. In short, due to the two reasons: a small dynamical critical region and the chemical freeze-out curves deviating from the phase boundary in the regime of large baryon chemical potential, the effects of critical slowing down are mild on the chemical freeze-out curves. Therefore, it is reasonable to expect that observables in heavy-ion collisions, which are mainly determined at the chemical freeze-out, are not strongly affected by the critical slowing down.

Furthermore, it is very interesting to extend the QCD-assisted relaxation dynamic model discussed here to include, e.g., the conservation of energy and momentum, nontrivial Poisson bracket, etc., as in the dynamic Model G or Model H. We will report the relevant progress in the future.

\emph{Acknowledgements.--}
We thank Jan M. Pawlowski, Fabian Rennecke, Johannes V. Roth, Lorenz von Smekal, Shanjin Wu for discussions and comments. We also would like to thank the members of the fQCD collaboration \cite{fQCD} for collaborations on related projects. This work is supported by the National Natural Science Foundation of China under Contract Nos.\ 12447102, 12175030. S. Yin and Y.-r. Chen are supported by Alexander v. Humboldt Foundation. C. Huang is supported by the Collaborative Research Centre SFB 1225 (ISOQUANT).

\bibliography{ref-lib}

\newpage

\renewcommand{\thesubsection}{{S.\arabic{subsection}}}
\setcounter{section}{0}
\titleformat*{\section}{\centering \Large \bfseries}

\onecolumngrid

\section*{Supplemental Materials}

The supplemental materials provide some details of inputs of Langevin transport model from the fRG calculations of QCD at finite temperature and densities, \Cref{app:inputs}, setup of Langevin simulations and robustness check, \Cref{app:Setup-robust}, some threshold functions, \Cref{app:thres-fun}, behavior of the retarded two-point correlation function for the sigma mode in the limit of zero spatial momentum, \Cref{app:behav-limit}.


\subsection{Inputs of Langevin transport model from the fRG calculations of QCD at finite temperature and densities}
\label{app:inputs}

It is more convenient to rewrite the Langevin equation in main text in terms of the renormalized field in numerical calculations, which reads
\begin{subequations} 
\label{eq:Langevin-transport-renor}
\begin{align}
  \bar Z_\phi^{(t)}\partial_t \bar\sigma-\partial_i^2 \bar \sigma+U'(\bar \sigma)=\xi\,,\label{eq:Langevin-Eq-renor}
\end{align}
with the Gaussian noise 
\begin{align}
  \langle \xi(t,\bm x)\xi(t',\bm x')\rangle=2\bar Z_\phi^{(t)}T\delta(t-t')\delta(\bm x-\bm x')\,.\label{eq:Langevin-correlation-renor}
\end{align}
\end{subequations}
Here the renormalized field is given by
\begin{align}
  \bar\sigma=\sqrt{Z^{(i)}_{\phi}}\sigma\,,\label{}
\end{align}
and the renormalized temporal wave function reads
\begin{align}
  \bar Z_\phi^{(t)}=\frac{Z_\phi^{(t)}}{Z^{(i)}_{\phi}}\,,\label{eq:barZt}
\end{align}
respectively. The renormalized temporal wave function represents the damping rate of the sigma mode.

\subsubsection{Damping rate}
\label{app:damping-rate}

Due to the progress in the studies of QCD in the vacuum and at finite temperature and densities within the functional renormalization group (fRG) approach in recent years \cite{Mitter:2014wpa, Braun:2014ata, Rennecke:2015eba, Cyrol:2016tym, Cyrol:2017ewj, Corell:2018yil, Fu:2019hdw, Braun:2020ada, Braun:2023qak, Ihssen:2024miv, Fu:2024rto, Fu:2025hcm}, cf. \cite{Dupuis:2020fhh, Fu:2022gou} for recent reviews, it is now possible to describe the damping property of the sigma mode of QCD matter quantitatively from QCD within the fRG approach. As we have discussed in the main text, the computation of the temporal wave function $Z_\phi^{(t)}$ relies on the retarded two-point correlation function of the sigma mode, which is described in \cite{Fu:2024rto}. The calculation of spatial wave function $Z^{(i)}_{\phi}$ is less involved, that can be found in \cite{Fu:2019hdw}.

\begin{figure}[t]
\includegraphics[width=0.7\textwidth]{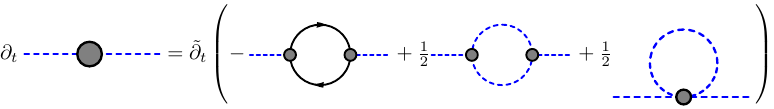}
\caption{Diagrammatic representation of the flow of two-point correlation functions for mesons, where the gray blobs denote the one-particle-irreducible (1PI) vertices. The derivative $\tilde \partial_t$ only hits the $k$-dependence of the regulators, which in fact results in an insert of one regulator for each inner propagator of the diagrams on the r.h.s. Here, $t=\ln(k/\Lambda)$ denotes the RG time, with $k$ and $\Lambda$ being the RG scale and an ultraviolet cutoff scale, respectively.}\label{fig:flowsGam2phiphi}
\end{figure}

Here we briefly recapitulate the method to calculate the real-time two-point correlation functions of mesons, e.g., the retarded two-point correlation function of sigma in the main text. The flow equation of two-point mesonic correlation functions is depicted in \Fig{fig:flowsGam2phiphi}, where the dashed line denotes the meson fields $\phi=(\sigma, \pi)$ including the sigma mode and pion, and the solid line stands for the quark field. One can see that there are three different parts in the flows, which are the quark loop, meson loop and the tadpole, respectively. Their explicit expressions for both the sigma and pion correlation functions can be found in \cite{Fu:2024rto}. The ingredients of the flow, i.e., the quark and meson propagators, Yukawa coupling and the mesonic vertices on the right side of the flow equation in \Fig{fig:flowsGam2phiphi}, are input from the calculations with the fRG approach to QCD at finite temperature and densities in \cite{Fu:2019hdw}. Note that the analytic continuation of correlation functions from the Euclidean to Minkowski spacetime is done on the level of the algebraic flow equation rather than on the Euclidean data of correlation functions. Therefore, there is no bothersome difficulty in ill-defined inverse problems, that is usually encountered when the Euclidean data have to be continued analytically into the regime of Minkowski spacetime on the level of numerical data. For more details about the calculations, we refer to Ref.~\cite{Fu:2024rto}.

%
%

%
\begin{figure*}[t]
\includegraphics[width=0.5\textwidth]{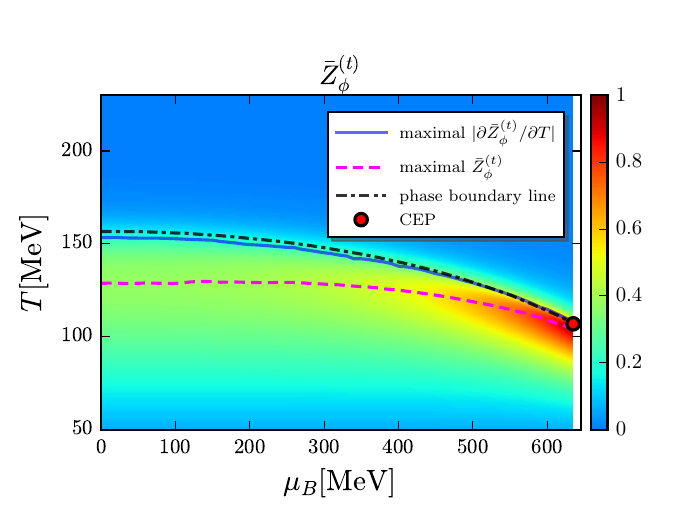}
\caption{Renormalized temporal wave function $\bar Z_\phi^{(t)}$ in \eq{eq:barZt} in the QCD phase diagram calculated from the fRG approach to QCD at finite temperature and densities, where $\bar Z_\phi^{(t)}$ is rescaled with the maximal value $\bar Z_{\phi, \mathrm{max}}^{(t)}$ in the phase diagram shown in this plot. The blue solid line denotes the position of the maximal value of $|\partial \bar Z_\phi^{(t)}/\partial T|$ along the temperature direction at different baryon chemical potentials. The magenta dashed line stands for the position of the maximal value of $\bar Z_\phi^{(t)}$ along the temperature direction at different baryon chemical potentials. The black dot-dashed line denotes the phase boundary.}\label{fig:Zt}
\end{figure*}
%

In \Fig{fig:Zt} we show the renormalized temporal wave function $\bar Z_\phi^{(t)}$ in \eq{eq:barZt} in the QCD phase diagram, calculated from QCD at finite temperature and densities within the fRG approach, whose setup is presented in \cite{Fu:2019hdw, Fu:2024rto}. Here, the value of $\bar Z_\phi^{(t)}$ has already been rescaled with the maximal value $\bar Z_{\phi, \mathrm{max}}^{(t)}$ in the phase diagram shown in this plot. The blue and magenta lines correspond to the maximal positions of $|\partial \bar Z_\phi^{(t)}/\partial T|$ and $\bar Z_\phi^{(t)}$ along the $T$ direction at a fixed $\mu_B$, respectively. The black dot-dashed line represents the phase boundary. One can see that with the increase of the baryon chemical potential, the three different curves approach towards each other, and finally converge at the critical end point (CEP), located at $(T_{_\mathrm{CEP}},\mu_{B_{\mathrm{CEP}}})=(107,635)$ MeV. There is a red region near the CEP, where $\bar Z_\phi^{(t)}$ increases remarkably. Note that the damping rate $\Gamma_D$ is inverse proportional to the renormalized temporal wave function \cite{Tan:2024fuq}, i.e., 
\begin{align}
    \Gamma_D \sim \frac{1}{\bar Z_\phi^{(t)}}=\frac{Z^{(i)}_{\phi}}{Z_\phi^{(t)}}\,,\label{}
\end{align}
which implies that the damping rate decreases significantly near the CEP. This accounts for part reason of the critical slowing down near the CEP. The other reason is due to the effective potential of the sigma field, which becomes very flat near the CEP, discussed in \Cref{app:potential}.

%
\begin{figure*}[t]
  \includegraphics[width=0.32\textwidth]{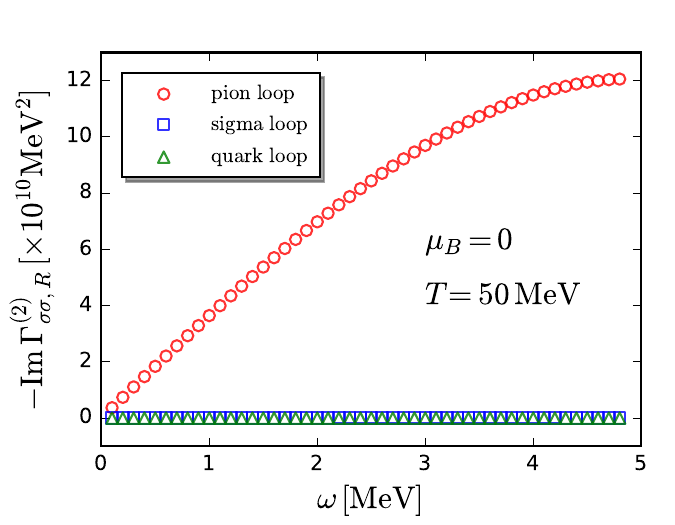} 
  \includegraphics[width=0.32\textwidth]{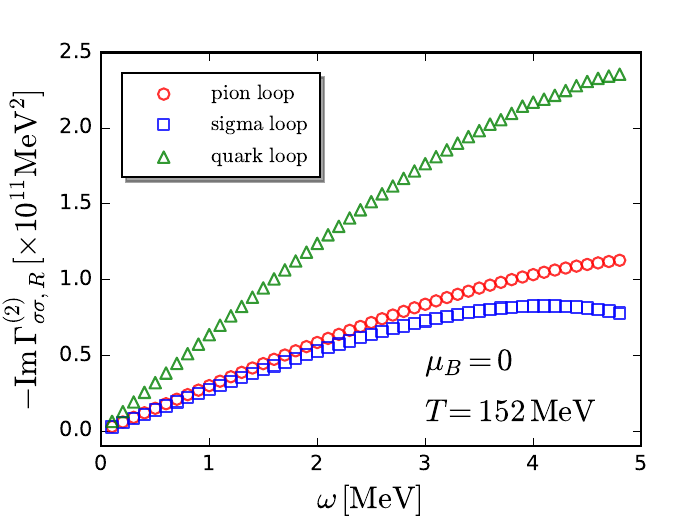}
  \includegraphics[width=0.32\textwidth]{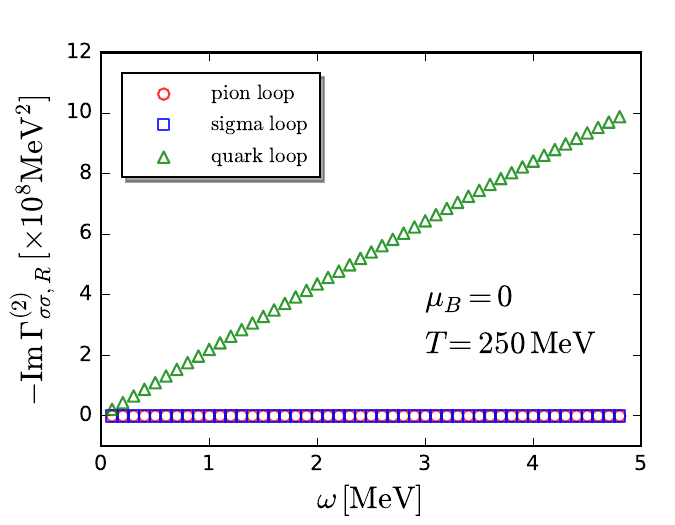}
  \includegraphics[width=0.32\textwidth]{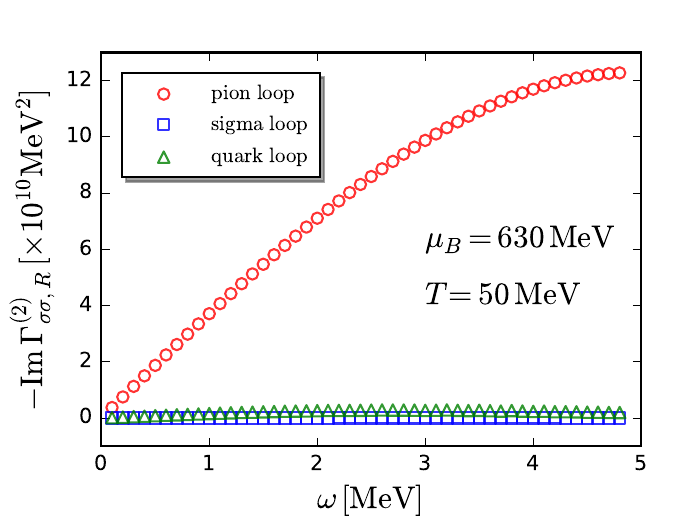}
  \includegraphics[width=0.32\textwidth]{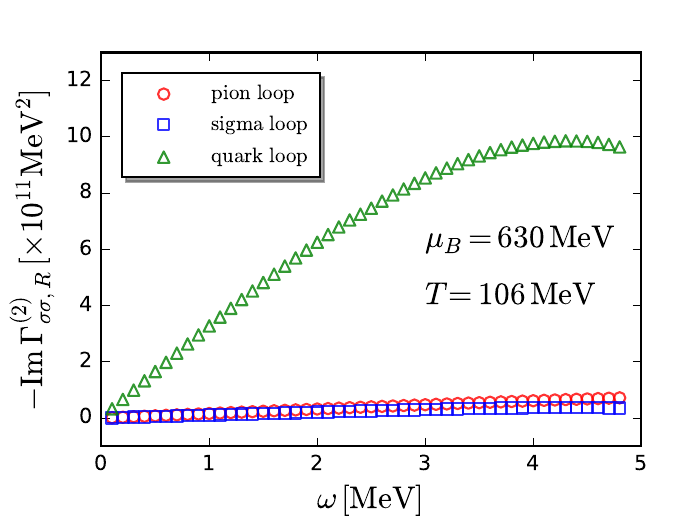}
  \includegraphics[width=0.32\textwidth]{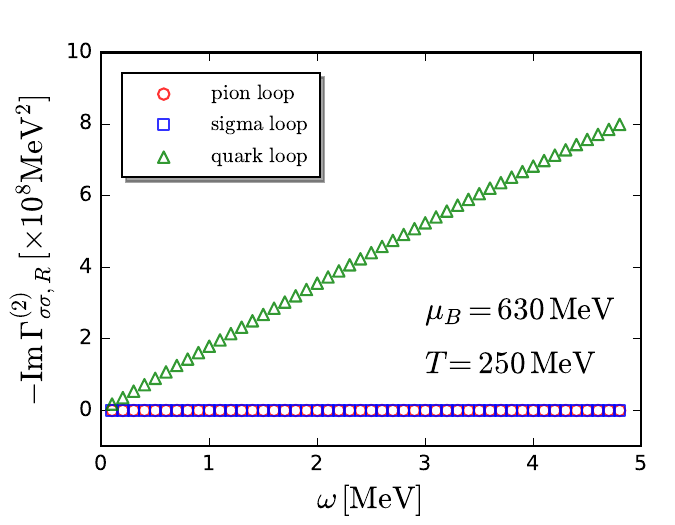}
\caption{Imaginary part of the retarded two-point correlation function of sigma mode as a function of the frequency, where contributions from different parts are presented separately. The spatial momentum is chosen to be $|\bm{p}|=10$ MeV. Results with $\mu_B=0$, 630 MeV for several different values of $T$ are presented in the panels of the first and second rows, respectively.}\label{fig:ImGamma}
\end{figure*}
%

In \Fig{fig:ImGamma} we show the imaginary part of the retarded two-point correlation function of sigma mode as a function of the frequency near $\omega \to 0$. From the slope of the curves at the origin, one is able to obtain the transport coefficient, i.e., the temporal wave function, as we have discussed in the main text. Different contributions from the quark, sigma, and the pion loops as shown on the right side in \Fig{fig:flowsGam2phiphi} are presented separately. Note that the last diagram of tadpole in \Fig{fig:flowsGam2phiphi} does not contribute to the imaginary part of the retarded correlation function. From \Fig{fig:ImGamma} one finds that the pion loop dominates in the regime of low temperature due to its Goldstone nature of chiral symmetry breaking. As the temperature increases, the quark loop begins to play a role and is comparable with the pion and sigma loops near the phase boundary, and eventually the quark loop plays a dominating role in the regime of high temperature, where the mesonic degrees of freedom decouple from the system. Note that the magnitude of the imaginary part of correlation functions arrives at its peak near the phase boundary.

\subsubsection{Effective potential}
\label{app:potential}

%
\begin{figure*}[t]
  \includegraphics[width=0.49\textwidth]{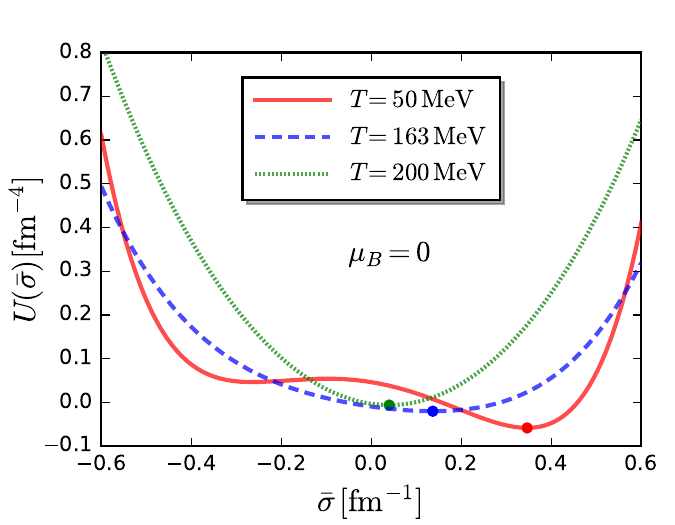} \hfill
  \includegraphics[width=0.49\textwidth]{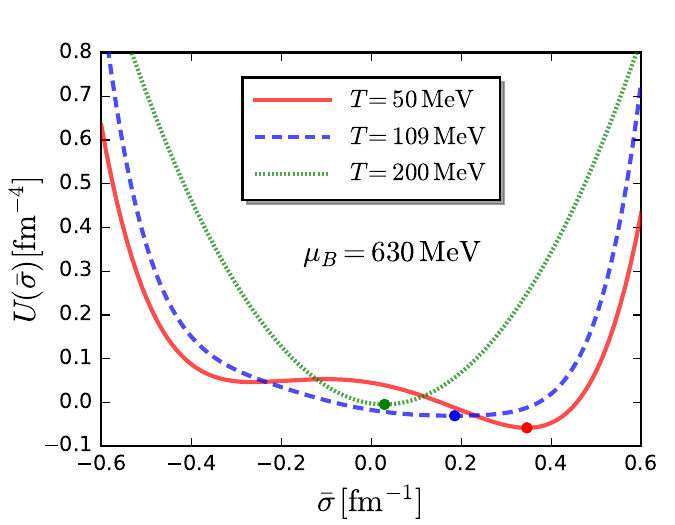}
\caption{Effective potential as a function of the renormalized sigma field for several different values of temperature at $\mu_B=0$ (left panel) and $\mu_B=630$ MeV (right panel), calculated from QCD at finite temperature and densities within the fRG approach \cite{Fu:2019hdw}. The dots on each curve indicate the minimum position of the effective potentials.}\label{fig:U-sigma}
\end{figure*}
%

%
\begin{figure*}[t]
  \includegraphics[width=0.49\textwidth]{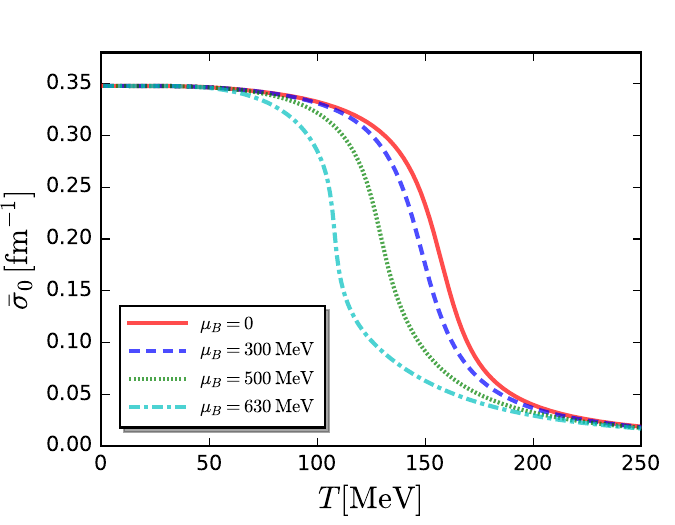} \hfill
  \includegraphics[width=0.49\textwidth]{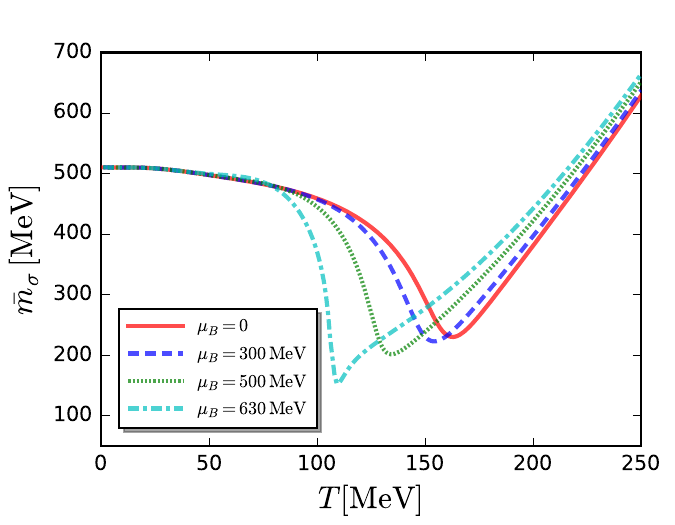}
\caption{\emph{Left panel}: Expectation value of the renormalized sigma field, i.e., $U'(\bar \sigma_0)=0$, as a function of the temperature for several different values of baryon chemical potential. \emph{Right panel}: Renormalized curvature mass of the sigma at the minimum of effective potential, i.e., $\bar m_{\sigma}=\sqrt{U^{(2)}(\bar \sigma_0)}$, as a function of the temperature for several different values of baryon chemical potential.}\label{fig:sigma0-msigma}
\end{figure*}
%

%
\begin{figure*}[t]
  \includegraphics[width=0.49\textwidth]{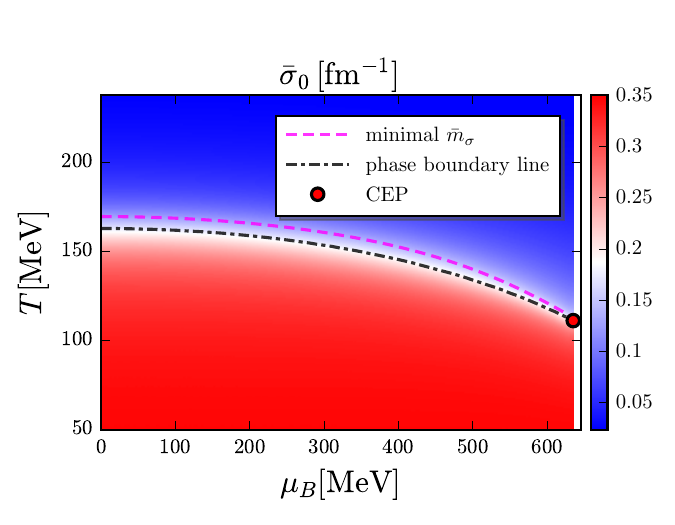} \hfill
  \includegraphics[width=0.49\textwidth]{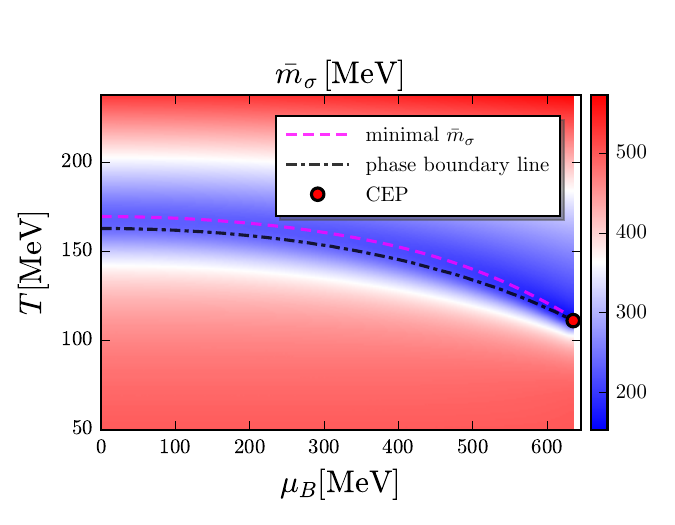}
\caption{Heatmaps of the expectation value of the renormalized sigma field (left panel) and the sigma curvature mass (right panel) in the QCD phase diagram. The magenta dashed line stands for the position of minimal $\bar m_{\sigma}$ along the temperature direction at different baryon chemical potentials. The black dot-dashed line denotes the phase boundary.}\label{fig:sigma0-msigma-phasedigram}
\end{figure*}
%

The effective potential in \Eq{eq:Langevin-transport-renor} is computed from the QCD at finite temperature and densities within the fRG \cite{Fu:2019hdw}, which is also discussed in \cite{Braun:2023qak}. In \Fig{fig:U-sigma} we show the effective potential as a function of the order parameter field $\bar \sigma$ for different temperature and baryon chemical potentials. The dots denote the position of the minimum of potentials with $\bar \sigma=\bar \sigma_0$, i.e., $U'(\bar \sigma_0)=0$. Here $\bar \sigma_0$ is also called the expectation value of the sigma field. The curvature of effective potential at $\bar \sigma=\bar \sigma_0$ is related to the curvature mass of sigma, i.e.,
\begin{align}
    \bar m_{\sigma}^2=U^{(2)}(\bar \sigma_0)\,,\label{}
\end{align}
which describes the flatness of the potential. Smaller is the curvature mass of sigma, flatter is the effective potential, which results in longer time for the relaxation from nonequilibrium to equilibrium. In \Fig{fig:sigma0-msigma} the expectation value of the sigma field $\bar \sigma_0$ and the curvature mass of sigma $\bar m_{\sigma}$ are presented as functions of the temperature for several values of baryon chemical potential. One can see that the minimum of the sigma mass decreases with the increase of $\mu_B$. In \Fig{fig:sigma0-msigma-phasedigram} $\bar \sigma_0$ and $\bar m_{\sigma}$ are depicted in the QCD phase diagram, where the phase boundary and the line of minimal $\bar m_{\sigma}$ along the temperature direction at fixed $\mu_B$ are also shown.

\subsection{Setup of Langevin simulations and robustness check}
\label{app:Setup-robust}

\subsubsection{Lattice spacing, size and ensemble number}

%
\begin{figure*}[t]
\includegraphics[width=0.5\textwidth]{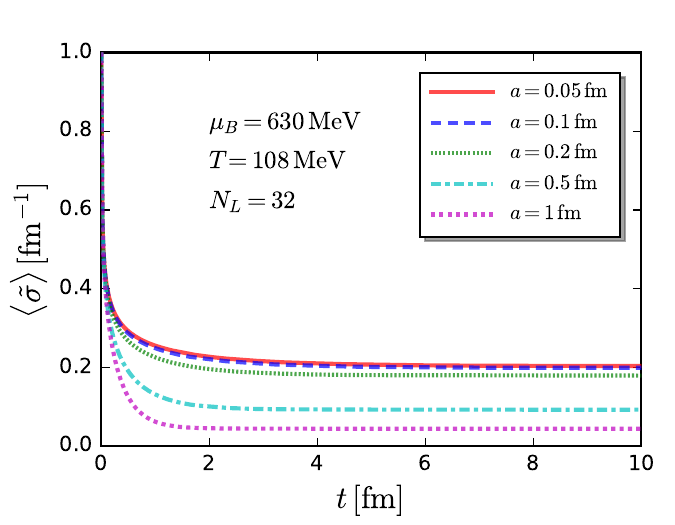}
\caption{Time evolution of ensemble average of the sigma field $\langle\tilde\sigma(t)\rangle$ for different lattice spacings, where the baryon chemical potential $\mu_B=630$ MeV and the temperature $T=108$ MeV are chosen. The lattice size is fixed to be $N_L=32$. The initial configuration of field is chosen to be homogeneous with $\bar \sigma(t=0,\bm x)=1\, \mathrm{fm^{-1}}$.}
\label{fig:lattice-spacing}
\end{figure*}
%

%
\begin{figure*}[t]
\includegraphics[width=0.45\textwidth]{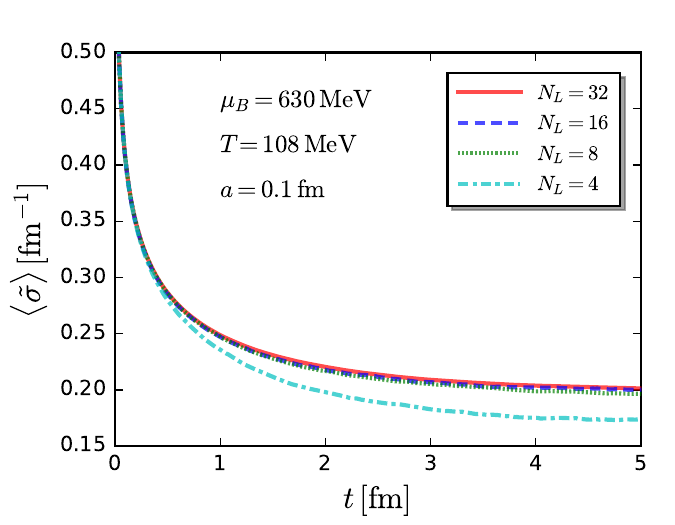}\hspace{0.4cm}
\includegraphics[width=0.45\textwidth]{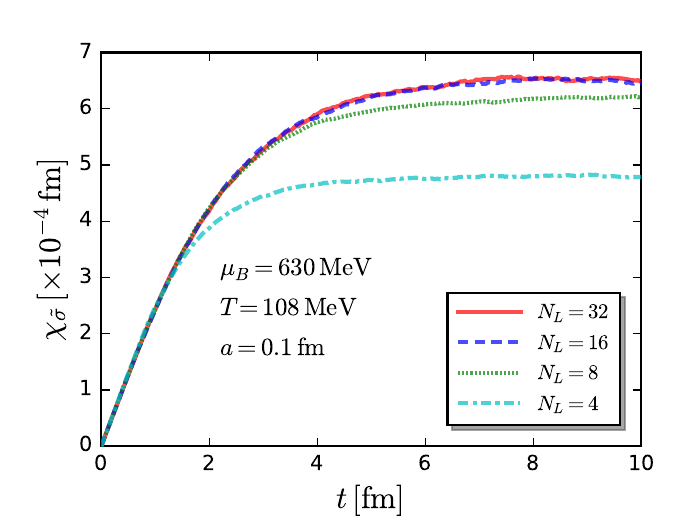}
\caption{Time evolution of ensemble average of the sigma field $\langle\tilde\sigma(t)\rangle$ (left panel) and the susceptibility $\chi_{\tilde\sigma}(t)$ (right panel) for different lattice sizes, where the baryon chemical potential $\mu_B=630$ MeV and the temperature $T=108$ MeV are chosen. The lattice spacing is fixed to be $a=0.1$ fm.}
\label{fig:lattice-number}
\end{figure*}
%

%
\begin{figure}[t]
\includegraphics[width=0.5\textwidth]{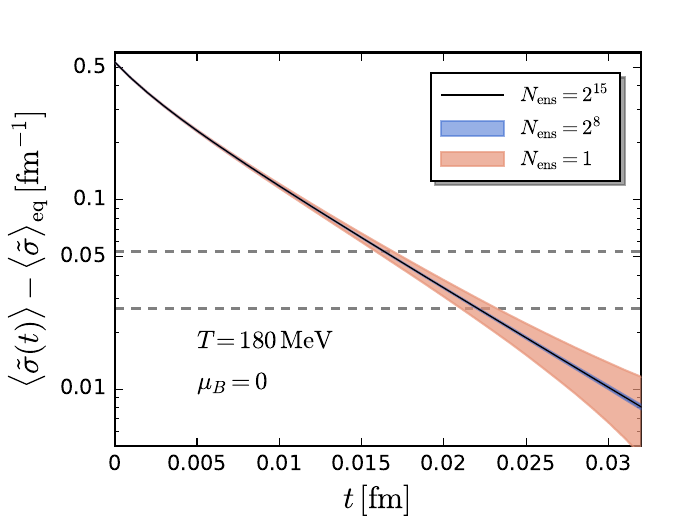}
\caption{Time evolution of the ensemble average of the sigma field subtracted by its equilibrium value, $\langle\tilde\sigma(t)\rangle- \langle\tilde\sigma\rangle_{\mathrm{eq}}$, for different sizes of the ensemble with $\mu_B=0$ and $T=180$ MeV. The colored bands represent the respective standard deviations. The region between the two horizontal dashed lines denotes the range used for the fitting of the relaxation time.}
\label{fig:ensemble-error}
\end{figure}
%

As we have discussed in the main text, we simulate the QCD-assisted Langevin equation in 3+1 dimensions. A three-dimensional periodic spatial lattice is used, which is characterized by its lattice spacing or lattice constant $a$ and lattice size $N_L^3$. The time step is chosen adaptively with the control of relative error. Moreover, since the evolution of the Langevin equation is sourced by stochastic noises, an ensemble of field configurations is required. The ensemble is characterized by the sample number $N_{\mathrm{ens}}$ in the ensemble. In the following we would like to investigate the dependence of simulations on the lattice constant $a$, lattice size $N_L$ and the ensemble number $N_{\mathrm{ens}}$.

In \Fig{fig:lattice-spacing} we show the time evolution of the sigma field simulated in lattices with different lattice spacings. The baryon chemical potential and temperature are chosen to be $\mu_B=630$ MeV and $T=108$ MeV, respectively. Its location in the QCD phase diagram is just inside the golden red region as shown in the left panel of Fig.~1 in the main text, being in the proximity of the CEP. It is found obviously in \Fig{fig:lattice-spacing} that a convergent result is obtained when the lattice spacing is decreased down to $a \sim 0.1$ fm, where the effect of finite lattice spacing is negligible.

In the same way, the effect of finite sizes of lattice is investigated in \Fig{fig:lattice-number}. Generally speaking, the lattice size should be larger than the correlation length of the critical mode such that the effect of finite size can be suppressed, that is particularly pronounced when the system is near the critical end point where the correlation length increases significantly. We choose the same point with $\mu_B=630$ MeV and $T=108$ MeV in the phase diagram as a test point near the CEP. The time evolution of the sigma field for different lattice sizes is depicted in the left panel of \Fig{fig:lattice-number}. One can see that convergent results are obtained when $N_L \gtrsim 8$ with $a=0.1$ fm. Moreover, we also calculate the susceptibility of the sigma field, i.e.,
\begin{align}
  \chi_{\tilde{\sigma}}=V\big(\langle\tilde{\sigma}^2\rangle-\langle\tilde{\sigma}\rangle^2\big)=a^3 N_L^3\big(\langle\tilde{\sigma}^2\rangle-\langle\tilde{\sigma}\rangle^2\big)\,,\label{eq:suscep}
\end{align}
which is a quantity of second-order fluctuations, and thus is more sensitive to the effect of finite sizes than the field itself. The relevant numerical results are presented in the right panel of \Fig{fig:lattice-number}. In the calculations of $\chi_{\tilde\sigma}$ in \eq{eq:suscep} the ensemble average is done over the samples of number from $N_{\mathrm{ens}}=2^{16}=65536$ to $2^{18}=262144$, which are large enough to suppress the statistical errors, see the discussions below. From the right panel of \Fig{fig:lattice-number}, one finds that the susceptibility is indeed more liable to the finite-size effect, but a convergence for the susceptibility is still observed when $N_L \gtrsim 16$ with $a=0.1$ fm.

In summary, we have studied the dependence of Langevin simulations on the lattice spacing and lattice size. It is found that simulations with lattice spacing $a=0.1$ fm and lattice size $N_L=32$ have arrived at convergent results. These lattice parameters are used throughout this work, whereby results in the main text are obtained. This choice also allows for the balance between the computational cost and the error minimization.

We also investigate the dependence of simulations on the size of ensemble, i.e., the sample number of the ensemble $N_{\mathrm{ens}}$. The time evolution of the ensemble average of the sigma field, subtracted by its equilibrium value, is shown in \Fig{fig:ensemble-error} for different sizes of the ensemble. Here we choose the vanishing baryon chemical potential and a relatively higher temperature $T=180$ MeV on purpose, since higher temperature implies more stochastic noises such that larger ensemble is required to suppress the statistical error. The colored bands represent the respective standard deviation for different sizes of ensemble. In fact, the error is reduced with the increase of the size of ensemble approximately through $\sim 1/\sqrt{N_{\mathrm{ens}}}$. It is found in \Fig{fig:ensemble-error} that the error is negligible as the size of ensemble is increased up to, say $N_{\mathrm{ens}}=2^{8}=256$.

\subsubsection{Initial condition dependence}

%
\begin{figure*}[t]
\includegraphics[width=0.45\textwidth]{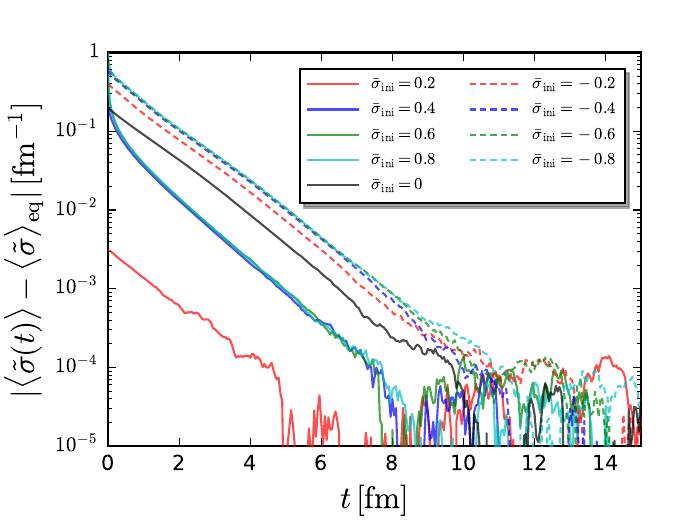}\hspace{0.4cm}
\includegraphics[width=0.45\textwidth]{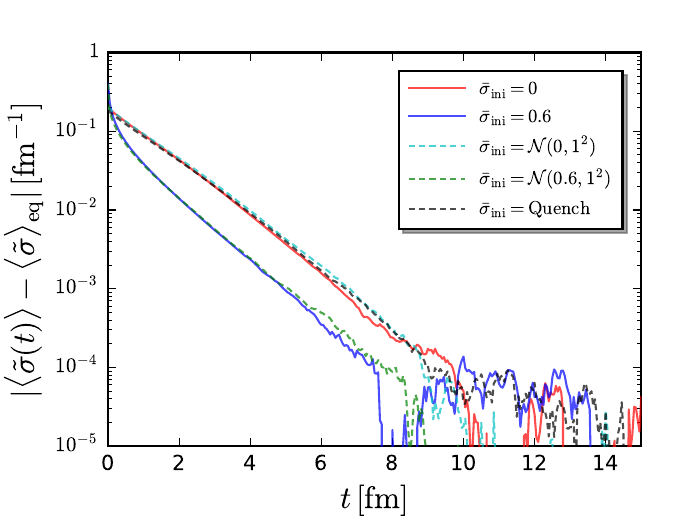}
\caption{Logarithmic-linear plots for the time evolution of the absolute value between the ensemble average of the sigma field and its equilibrium value for different initial conditions, where $\mu_B=630$ MeV and $T=108$ MeV are chosen. In the left panel, a constant initial field configuration $\bar{\sigma}(t=0,\bm{x})=\bar{\sigma}_{\mathrm{ini}}$ with different values of $\bar{\sigma}_{\mathrm{ini}}$ is used. In the right panel, the initial field configuration is also extend to the cases of the Gaussian distribution and the quenched initial condition.}
\label{fig:time-evolution_ini}
\end{figure*}
%

As we have discussed in the main text that the relaxation time is insensitive to the initial field configurations. To demonstrate this point, we perform the Langevin simulations with different initial conditions. The results are presented in \Fig{fig:time-evolution_ini}. In the left panel, a constant initial field configuration $\bar{\sigma}(t=0,\bm{x})=\bar{\sigma}_{\mathrm{ini}}$ is employed, with $\bar{\sigma}_{\mathrm{ini}}$ ranging from $-0.8\, \mathrm{fm}^{-1}$ to $0.8\, \mathrm{fm}^{-1}$. We plot the $\left|\langle\tilde\sigma(t)\rangle-\langle\tilde\sigma\rangle_\mathrm{eq}\right|$ in terms of logarithmic coordinate for the $y$-axis, such that one immediately recognizes that the slope of curves is inversely proportional to the relaxation time. We find that all these curves with different values of constant initial field are parallel to each other in their medium stage of evolution, indicating that they are characterized by the same relaxation time. In the right panel of \Fig{fig:time-evolution_ini}, the initial configuration is also extended to other cases, such as the Gaussian distribution, i.e., $\bar{\sigma}(t=0,\bm{x})\sim\mathcal{N}(\mu,\sigma_{\mu}^2)$ with the expectation value $\mu$ and standard deviation $\sigma_{\mu}$, and the quenched initial condition, where the initial ensemble is prepared with the equilibrium distribution at a higher temperature $T=250$ MeV and the same $\mu_B=630$ MeV. One can see that they have the same relaxation time as the constant initial configurations. In short, the relaxation time of critical mode is insensitive to the initial conditions of different types.

\subsubsection{Transport coefficient dependence}

%
\begin{figure}[t]
\includegraphics[width=0.5\textwidth]{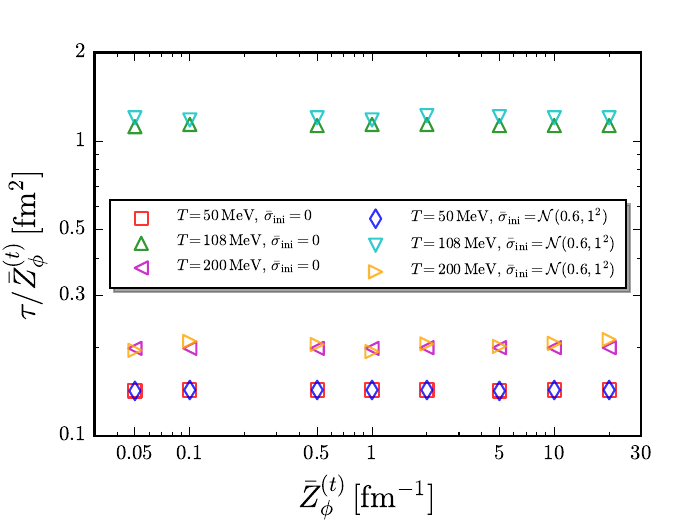}
\caption{Relaxation time divided by the transport coefficient, i.e., the renormalized temporal wave function $\bar Z_\phi^{(t)}$, as a function of the transport coefficient. Three different values of temperature are chosen at a fixed baryon chemical potential with $\mu_B=630$ MeV. Constant and Gaussian initial field configurations are used.}
\label{fig:relaxtime_Zt}
\end{figure}
%

From \Eq{eq:Langevin-transport-renor}, one readily arrives at the damping frequency
\begin{align}
    \omega\sim \frac{\bar m_{\sigma}^2}{\bar Z_\phi^{(t)}}\,,\label{}
\end{align}
in the limit of vanishing momentum $\bm{p} \to 0 $ \cite{Tan:2024fuq}. Considering the relaxation time $\tau \sim 1/\omega$, one is led to
\begin{align}
    \tau \sim \frac{\bar Z_\phi^{(t)}}{\bar m_{\sigma}^2}\,.\label{eq:tau-Zt-msig2}
\end{align}
In \Fig{fig:relaxtime_Zt} we investigate the dependence of the ratio $\tau/\bar Z_\phi^{(t)}$ on the transport coefficient, i.e., the renormalized temporal wave function $\bar Z_\phi^{(t)}$, where several different values of temperature and different initial field configurations are employed. One can see that the ratio $\tau/\bar Z_\phi^{(t)}$ is almost independent of the transport coefficient for each value of temperature, and this is also insensitive the choice of initial field configurations. This indicates that the linear relation between $\tau$ and $\bar Z_\phi^{(t)}$ in \Eq{eq:tau-Zt-msig2} is fulfilled in the simulations.

\subsection{Threshold functions}
\label{app:thres-fun}

In the calculation of the quark loop appearing in the flow of two-point functions for mesons as shown in \Fig{fig:flowsGam2phiphi}, we need the threshold function defined as follows
\begin{align}
      &\mathcal{FF}_{(m,n)}(p)\equiv k^{2(m+n)-1}\,T\sum_{n_q}\Big(\bar G_q(q, \bar m_1^2)\Big)^m \Big(\bar G_q(q- p, \bar m_2^2)\Big)^n\,,\label{eq:FFmn}
\end{align}
where the Matsubara frequency is summed for $m$ fermionic propagators of mass $\bar{m}_1$ and $n$ fermionic propagators of mass $\bar{m}_2$. The quark propagator reads
\begin{align}
    \bar G_{q}(q)&=\frac{1}{(q_0+i \mu)^2+E_k^2(q,\bar m^2)} \,,\label{eq: barGq2}
\end{align}
with
\begin{align}
    E_k(q,\bar m^2)&=\sqrt{\bm{q}^2\big(1+r_F(\bm{q}^2/k^2)\big)^2+\bar m^2}\,,
\end{align}
where we have used the $3d$ optimized infrared regulator for the quark \cite{Litim:2000ci, Litim:2001up}
\begin{align}
    r_{F}(x)=\left(\frac{1}{\sqrt{x}}-1\right)\Theta(1-x)\,,\label{eq:rFopt}
\end{align}
with the Heaviside step function $\Theta(x)$.

As we have discussed in the main text, in order to calculate the temporal wave function, one has to perform the analytic continuation for the two-point correlation function of mesons from the Euclidean to Minkowski spacetime. Thus, it is convenient to define some sort of functions, such as
\begin{align}
    I_k(p; \bar{m}_1, \bar{m}_2; \mathcal{FF}_{(m,n)})\equiv \int \frac{d^3\bm{q}}{(2\pi)^3}(\cdots) \mathcal{FF}_{(m,n)}\big(p_0\to -\mathrm{i}(p_0+\mathrm{i}\epsilon),\bm{p}\big)\,,\label{eq:Ik}
\end{align}
where $(\cdots)$ denotes some function being trivial with the analytic continuation, and the integral over the spatial momentum is performed. Here, we focus on the lowest-order threshold function $\mathcal{FF}_{(1,1)}$, and use the abbreviation as follows
\begin{align}
    I_k(p; \bar{m}_1, \bar{m}_2)=I_k(p; \bar{m}_1, \bar{m}_2; \mathcal{FF}_{(1,1)})\,.\label{}
\end{align}
The higher-order ones with $\mathcal{FF}_{(m,n)}$ can be obtained by taking the derivative of the $I_k$ function with respect to $\bar{m}_1$ or/and $\bar{m}_2$. The subscript $k$ indicates the renormalization group (RG) dependence of the $I_k$ function.

It is convenient to express $I_k$ as a sum of two parts, that is,
\begin{align}
    I_k^f(p;\bar{m}_1,\bar{m}_2)&=I_{1,k}^f(p;\bar{m}_1,\bar{m}_2)+I_{2,k}^f(p;\bar{m}_1,\bar{m}_2)\,, \label{eq:I1+I2}
\end{align}
where we have added a superscript $f$ to indicate that the $I_k$ function depends on some function $f$ arising from $(\cdots)$ in \Eq{eq:Ik}, also cf. \Cref{eq:ImI1,eq:ImI2}. On the right side of \Eq{eq:I1+I2} $I_1$ and $I_2$ stand for the contributions from the creation-annihilation process of quark and anti-quark pair and the Landau damping (particle-hole process of quark near the Fermi surface), respectively. The two different processes are analyzed in detail in \cite{Fu:2024rto}. We are interested in the imaginary parts of $I_1$ and $I_2$, since they are related to the temporal wave function discussed in the main text. Inserting \Eq{eq:FFmn} into \Eq{eq:Ik}, one arrives at
\begin{align}
    &\mathrm{Im} I_{1,k}^f(p\,;\bar{m}_1,\bar{m}_2)\nonumber\\[2ex]
    =&\int_{0}^{k}d|\bm{q}|\int_{-1}^{1}{d\cos\theta}\;\;f(|\bm{q}|,\cos\theta)\bigg[\delta\Big(E_{k}(q,\bar{m}_1^2)+E_{k}(q-p,\bar{m}_2^2)-p^0\Big)-\delta\Big(-E_{k}(q,\bar{m}_1^2)-E_{k}(q-p,\bar{m}_2^2)-p^0\Big)\bigg] \,, \label{eq:ImI1} \\[2ex]
    &\mathrm{Im} I_{2,k}^f(p\,;\bar{m}_1,\bar{m}_2)\nonumber\\[2ex]
    =&\int_{0}^{k}d|\bm{q}|\int_{-1}^{1}{d\cos\theta}\;\;f(|\bm{q}|,\cos\theta)\bigg[\delta\Big(-E_{k}(q,\bar{m}_1^2)+E_{k}(q-p,\bar{m}_2^2)-p^0\Big)-\delta\Big(E_{k}(q,\bar{m}_1^2)-E_{k}(q-p,\bar{m}_2^2)-p^0\Big)\bigg] \,, \label{eq:ImI2}
\end{align}
with some function $f(|\bm{q}|,\cos\theta)$ in the integrand. 

In order to show the results more concisely, we also define several other functions. The first one appearing in the $I_1$ function reads
\begin{align}
    \mathcal{F}^f_1(q_-,k,p;\bar{m}_1^2,\bar{m}_2^2)\equiv&\;\int_{q_-}^{k}dq\;\;\frac{p_0-E_k(\bar{m}_1^2)}{pq}f\left(q,\frac{p^2+q^2-q_-^2(k)}{2pq}\right)\,,
\end{align}
with
\begin{equation}
    q_-(q)\equiv\sqrt{\big(p_0-E(q,\bar{m}_1^2)\big)^2-\bar{m}_2^2}\,,
\end{equation}
where one has
\begin{align}
    E(x,\bar{m}^2)&\equiv \sqrt{x^2+\bar{m}^2}\,,\nonumber\\[2ex]
  E_k(\bar{m}^2)&\equiv E(k,\bar{m}^2)\,.  \label{}
\end{align}
Moreover, we also need another two functions proportional to the delta function, which read
\begin{align}
    &\mathcal{F}_2^f(p_0,k,p;\bar{m}_1^2,\bar{m}_2^2)\nonumber\\[2ex]
 \equiv&\;\delta\left(p_0-E_k(\bar{m}_1^2)-E_k(\bar{m}_2^2)\right)\Bigg\{\int_{0}^{k-p}\!dq\int_{-1}^{1}\!d\cos\theta\;f(q,\cos\theta)+\int_{k-p}^{k}\!dq\int_{\frac{p^2+q^2-k^2}{2pq}}^{1}\!d\cos\theta\;f(q,\cos\theta)\Bigg\}\,,\label{}
\end{align}
and
\begin{align}
    \mathcal{F}_3^f(p_0,k,p;\bar{m}_1^2,\bar{m}_2^2)&\equiv \delta\left(p_0-E_k(\bar{m}_1^2)-E_k(\bar{m}_2^2)\right)\int_{p-k}^{k}\!dq\int_{\frac{p^2+q^2-k^2}{2pq}}^{1}\!d\cos\theta\;f(q,\cos\theta)\,.\label{}
\end{align}
The function appearing in $I_2$ reads
\begin{align}
&\mathcal{F}_1'^f(q_-,k,p;\bar{m}_1^2,\bar{m}_2^2)\nonumber\\[2ex]
\equiv&\;\int_{q_-}^{k}dq\;\;\frac{p_0+E_k(\bar{m}_1^2)}{pq} f\left(q,\frac{p^2+q^2-q^2_+(k)}{2pq}\right)-\Big(\text{vacuum contributions}\Big)\,,\label{eq:F1pri}
\end{align}
with
\begin{align}
    q_+(q)\equiv\sqrt{\big(p_0+E(q,\bar{m}_1^2)\big)^2-\bar{m}_2^2}\,.
\end{align}

Now, it is ready to show the explicit expressions of $\mathrm{Im} I_{1,k}^f(p;\bar{m}_1,\bar{m}_2)$ and $\mathrm{Im} I_{2,k}^f(p;\bar{m}_1,\bar{m}_2)$ by means of piecewise functions. We begin with $\mathrm{Im} I_{1,k}^f(p;\bar{m}_1,\bar{m}_2)$.
\begin{enumerate}[I.]
\item If $k>|\bm p|$, one has
\begin{enumerate}[(1).]  
\item when $p^0\geq E_k(k,\bar{m}_1^2)+E_k(k+|\bm p|,\bar{m}_2^2)$, then
\begin{align}
  \mathrm{Im} I_{1,k}^f(p;\bar{m}_1,\bar{m}_2)=&0\,,  \label{}
\end{align}
\item when $p^0<E_k(k+|\bm p|,\bar{m}_1^2)+E_k(k,\bar{m}_2^2)$ and $p^0> E_k(\bar{m}_1^2)+E_k(\bar{m}_2^2)$, then
\begin{align}
  \mathrm{Im} I_{1,k}^f(p;\bar{m}_1,\bar{m}_2)=&\mathcal{F}_1^f(q_-,k,|\bm p|;\bar{m}_1^2,\bar{m}_2^2)\,,  \label{}
\end{align}
with
\begin{align}
  q_-&=\sqrt{\big(p^0-E(k,\bar{m}_1^2)\big)^2-\bar{m}_2^2}-|\bm p|\,;  \label{}
\end{align}
\item when $p^0=E_k(\bar{m}_1^2)+E_k(\bar{m}_2^2)$, then
\begin{align}
  \mathrm{Im} I_{1,k}^f(p;\bar{m}_1,\bar{m}_2)=&\mathcal{F}_2^f(p_0,k,|\bm p|;\bar{m}_1,\bar{m}_2)\,;  \label{}
\end{align}
\item when $p^0<E_k(\bar{m}_1^2)+E_k(\bar{m}_2^2)$, then
\begin{align}
  \mathrm{Im} I_{1,k}^f(p;\bar{m}_1,\bar{m}_2)=&0\,.  \label{}
\end{align}
\end{enumerate}
\item If $|\bm p|/2<k\leq |\bm p|$, one has
\begin{enumerate}[(1).]  
\item when $p^0\geq E(k+|\bm p|,\bar{m}_1^2)+E(k,\bar{m}_2^2)$, then
\begin{align}
  \mathrm{Im} I_{1,k}^f(p;\bar{m}_1,\bar{m}_2)=&0\,,  \label{}
\end{align}
\item when $p^0<E(k+|\bm p|,\bar{m}_1^2)+E(k,\bar{m}_2^2)$ and $p^0\geq \big(E(|\bm p|,\bar{m}_1^2)+E(k,\bar{m}_2^2)\big)$, then
\begin{align}
  \mathrm{Im} I_{1,k}^f(p;\bar{m}_1,\bar{m}_2)=&\mathcal{F}_1^f(q_-,k,|\bm p|;\bar{m}_1^2,\bar{m}_2^2)\,,  \label{}
\end{align}
with
\begin{align}
  q_-&=\sqrt{\big(p^0-E(k,\bar{m}_1^2)\big)^2-\bar{m}_2^2}-|\bm p|\,;  \label{}
\end{align}
\item when $p^0<E(|\bm p|,\bar{m}_1^2)+E(k,\bar{m}_2^2)$ and $p^0> E(k,\bar{m}_1^2)+E(k,\bar{m}_2^2)$, then
\begin{align}
  \mathrm{Im} I_{1,k}^f(p;\bar{m}_1,\bar{m}_2)=&\mathcal{F}_1^f(q_-,k,|\bm p|;\bar{m}_1^2,\bar{m}_2^2)\,,  \label{}
\end{align}
with
\begin{align}
  q_-&=-\sqrt{\big(p^0-E(k,\bar{m}_1^2)\big)^2-\bar{m}_2^2}+|\bm p|\,;  \label{}
\end{align}
\item when $p^0=E_k(\bar{m}_1^2)+E_k(\bar{m}_2^2)$, then
\begin{align}
  \mathrm{Im} I_{1,k}^f(p;\bar{m}_1,\bar{m}_2)=&\mathcal{F}_3^f(p_0,k,|\bm p|;\bar{m}_1,\bar{m}_2)\,;  \label{}
\end{align}
\item when $p^0<E_k(\bar{m}_1^2)+E_k(\bar{m}_2^2)$, then
\begin{align}
  \mathrm{Im} I_{1,k}^f(p;\bar{m}_1,\bar{m}_2)=&0\,.  \label{}
\end{align}
\end{enumerate}
\item If $k\leq |\bm p|/2$, one has
\begin{enumerate}[(1).]  
\item when $p^0\geq E(k+|\bm p|,\bar{m}_1^2)+E(k,\bar{m}_2^2)$, then
\begin{align}
  \mathrm{Im} I_{1,k}^f(p;\bar{m}_1,\bar{m}_2)=&0\,,  \label{}
\end{align}

\item when $p^0<E(k+|\bm p|,\bar{m}_1^2)+E(k,\bar{m}_2^2)$ and $p^0\geq E(|\bm p|,\bar{m}_1^2)+E(k,\bar{m}_2^2)$, then
\begin{align}
  \mathrm{Im} I_{1,k}^f(p;\bar{m}_1,\bar{m}_2)=&\mathcal{F}_1^f(q_-,k,|\bm p|;\bar{m}_1^2,\bar{m}_2^2)\,,  \label{}
\end{align}
with
\begin{align}
  q_-&=\sqrt{\big(p^0-E(k,\bar{m}_1^2)\big)^2-\bar{m}_2^2}-|\bm p|\,;  \label{}
\end{align}
\item when $p^0<E(|\bm p|,\bar{m}_1^2)+E(k,\bar{m}_2^2)$ and $p^0\geq E(k,\bar{m}_1^2)+E(k,\bar{m}_2^2)$, then
\begin{align}
  \mathrm{Im} I_{1,k}^f(p;\bar{m}_1,\bar{m}_2)=&\mathcal{F}_1^f(q_-,k,|\bm p|;\bar{m}_1^2,\bar{m}_2^2)\,,  \label{}
\end{align}
with
\begin{align}
  q_-&=-\sqrt{\big(p^0-E(k,\bar{m}_1^2)\big)^2-\bar{m}_2^2}+|\bm p|\,;  \label{}
\end{align}
\item when $p^0<E(k,\bar{m}_2^2)+E(|\bm p|-k,\bar{m}_1^2)$ and $p^0\geq E(|\bm p|/2,\bar{m}_1^2)+E(|\bm p|/2,\bar{m}_2^2)$, then
\begin{align}
  \mathrm{Im} I_{1,k}^f(p;\bar{m}_1,\bar{m}_2)=&0\,,  \label{}
\end{align}
\item when $p^0<E(|\bm p|/2,\bar{m}_1^2)+E(|\bm p|/2,\bar{m}_2^2)$, then
\begin{align}
  \mathrm{Im} I_{1,k}^f(p;\bar{m}_1,\bar{m}_2)=&0\,. \label{}\\
  \nonumber
\end{align}

\end{enumerate}
\end{enumerate}

Then, the expression of $\mathrm{Im} I_{2,k}^f(p;\bar{m}_1,\bar{m}_2)$ are given as follows. 
\begin{enumerate}[I.]
\item If $k>|\bm p|$, one has
\begin{enumerate}[(1).]  
\item when $p^0>|\bm p|$, then
\begin{align}
  \mathrm{Im} I_{2,k}^f(p;\bar{m}_1,\bar{m}_2)=&0\,,  \label{}
\end{align}
\item when $p^0>E(k+|\bm p|,\bar{m}_1^2)-E(k,\bar{m}_2^2)$ and $p^0\leq |\bm p|$, then
\begin{align}
  \mathrm{Im} I_{2,k}^f(p;\bar{m}_1,\bar{m}_2)=&0\,,  \label{}
\end{align}
\item when $p^0\leq E(k+|\bm p|,\bar{m}_1^2)-E(k,\bar{m}_2^2)$, then
\begin{align}
  \mathrm{Im} I_{2,k}^f(p;\bar{m}_1,\bar{m}_2)=&\mathcal{F}_1'^f(q_-,k,|\bm p|;\bar{m}_1^2,\bar{m}_2^2)\,,  \label{}
\end{align}
with
\begin{align}
  q_-&=\sqrt{\left(p^0+E(k,\bar{m}_1^2)\right)^2-\bar{m}_2^2}-|\bm p|\,.  \label{}
\end{align}
\end{enumerate}
\item If $|\bm p|/2<k\leq |\bm p|$, one has
\begin{enumerate}[(1).]  
\item when $p^0>|\bm p|$, then
\begin{align}
  \mathrm{Im} I_{2,k}^f(p;\bar{m}_1,\bar{m}_2)=&0\,,  \label{}
\end{align}
\item when $p^0>E(k+|\bm p|,\bar{m}_1^2)-E(k,\bar{m}_2^2)$ and $p^0\leq |\bm p|$, then
\begin{align}
  \mathrm{Im} I_{2,k}^f(p;\bar{m}_1,\bar{m}_2)=&0\,,  \label{}
\end{align}
\item when $p^0>E(|\bm p|,\bar{m}_1^2)-E(k,\bar{m}_2^2)$ and $p^0\leq E(k+|\bm p|,\bar{m}_1^2)-E(k,\bar{m}_2^2)$, then
\begin{align}
  \mathrm{Im} I_{2,k}^f(p;\bar{m}_1,\bar{m}_2)=&\mathcal{F}_1'^f(q_-,k,|\bm p|;\bar{m}_1^2,\bar{m}_2^2)\,,  \label{}
\end{align}
with
\begin{align}
  q_-&=\sqrt{\left(p^0+E(k,\bar{m}_1^2)\right)^2-\bar{m}_2^2}-|\bm p|\,,  \label{}
\end{align}
\item when $p^0\leq E(|\bm p|,\bar{m}_1^2)-E(k,\bar{m}_2^2)$, then
\begin{align}
  \mathrm{Im} I_{2,k}^f(p;\bar{m}_1,\bar{m}_2)=&\mathcal{F}_1'^f(q_-,k,|\bm p|;\bar{m}_1^{2},\bar{m}_2^2)\,,  \label{}
\end{align}
with
\begin{align}
  q_-&=-\sqrt{\left(p^0+E(k,\bar{m}_1^2)\right)^2-\bar{m}_2^{2}}+|\bm p|\,.  \label{}
\end{align}
\end{enumerate}
\item If $k\leq |\bm p|/2$, one has
\begin{enumerate}[(1).]  
\item when $p^0>|\bm p|$, then
\begin{align}
  \mathrm{Im} I_{2,k}^f(p;\bar{m}_1,\bar{m}_2)=&0\,,  \label{}
\end{align}
\item when $p^0>E(k+|\bm p|,\bar{m}_1^2)-E(k,\bar{m}_2^2)$ and $p^0\leq |\bm p|$, then
\begin{align}
  \mathrm{Im} I_{2,k}^f(p;\bar{m}_1,\bar{m}_2)=&0\,,  \label{}
\end{align}
\item when $p^0>E(|\bm p|,\bar{m}_1^2)-E(k,\bar{m}_2^2)$ and $p^0\leq E(k+|\bm p|,\bar{m}_1^2)-E(k,\bar{m}_2^2)$, then
\begin{align}
  \mathrm{Im} I_{2,k}^f(p;\bar{m}_1,\bar{m}_2)=&\mathcal{F}_1'^f(q_-,k,|\bm p|;\bar{m}_1^2,\bar{m}_2^2)\,,  \label{}
\end{align}
with
\begin{align}
  q_-&=\sqrt{\left(p^0+E(k,\bar{m}_1^2)\right)^2-\bar{m}_2^2}-|\bm p|\,,  \label{}
\end{align}
\item when $p^0>E(k-|\bm p|,\bar{m}_1^2)-E(k,\bar{m}_2^2)$ and $p^0\leq E(|\bm p|,\bar{m}_1^2)-E(k,\bar{m}_2^2)$, then
\begin{align}
  \mathrm{Im} I_{2,k}^f(p;\bar{m}_1,\bar{m}_2)=&\mathcal{F}_1'^f(q_-,k,|\bm p|;\bar{m}_1^2,\bar{m}_2^2)\,,  \label{}
\end{align}
with
\begin{align}
  q_-&=-\sqrt{\left(p^0+E(k,\bar{m}_1^2)\right)^2-\bar{m}_2^2}+|\bm p|\,,  \label{}
\end{align}
\item when $p^0\leq E(k-|\bm p|,\bar{m}_1^2)-E(k,\bar{m}_2^2)$, then
\begin{align}
  \mathrm{Im} I_{2,k}^f(p;\bar{m}_1,\bar{m}_2)=&0\,.  \label{}
\end{align}

\end{enumerate}
\end{enumerate}

\subsection{Behavior of the retarded two-point correlation function for the sigma mode in the limit of zero spatial momentum}
\label{app:behav-limit}

The retarded two-point correlation function for the sigma mode is related to the function $I_k(p; \bar{m}_1, \bar{m}_2; \mathcal{FF}_{(2,1)})$ in \Eq{eq:Ik} discussed in \Cref{app:thres-fun}. Here, we focus on the Landau damping part $I_{2,k}$ as shown in \Eq{eq:I1+I2}. We begin with the counterpart related to the threshold function $\mathcal{FF}_{(1,1)}$, whose imaginary part reads
\begin{align}
  &\mathrm{Im} I_{2,k}(p\,;\bar{m}_1,\bar{m}_2;\mathcal{FF}_{(1,1)})\nonumber\\[2ex]
  =&\int_{0}^{k}dq\int_{-1}^{1}{d\cos\theta}\;\;q^2\frac{n_F\left(E_k(q-p,\bar{m}_2^2)\right)-n_F(E_k(\bar{m}_1^2))}{4E_k(\bar{m}_1^2)E_k(q-p,\bar{m}_2^2)}\nonumber\\[2ex]
  &\times\bigg[\delta\Big(-E_{k}(q,\bar{m}_1^2)+E_{k}(q-p,\bar{m}_2^2)-p^0\Big)-\delta\Big(E_{k}(q,\bar{m}_1^2)-E_{k}(q-p,\bar{m}_2^2)-p^0\Big)\bigg] \,, 
\end{align}
with the fermionic distribution function
\begin{align}
  n_F(E)&=\frac{1}{\exp\Big[(E-\mu)/T\Big]+1}\,,
\end{align} 
where $T$ and $\mu$ are the temperature and quark chemical potential, respectively. Note that in numerical calculations the modification of the distribution function arising from the Polyakov loops are also taken into account, see, e.g., Eq.~(N8) in \cite{Fu:2019hdw}. Consequently, one finds for the function $\mathcal{F}_1'$ in \Eq{eq:F1pri}
\begin{align}
  \mathcal{F}_1'(q_-,k,|\bm p|;\bar{m}_1^2,\bar{m}_2^2)=&(k^2-q_-^2)\frac{n_F\left(E_k(\bar{m}_1^2)\right)-n_F\left(p_0+E_k(\bar{m}_1^2)\right)}{8E_k(\bar{m}_1^2)|\bm{p}|}\,,  \label{eq:F1pri2}
\end{align}
with
\begin{align}
  q_-&=\sqrt{\big(p^0-E(k,\bar{m}_1^2)\big)^2-\bar{m}_2^2}-|\bm p|\,.  \label{}
\end{align}
Expanding \Eq{eq:F1pri2} in powers of $p^0$ to the leading order, and using $\bar{m}_1=\bar{m}_2=\bar{m}_f$, one arrives at $q_-=k-|\bm p|$ and $n_F\left(E_k(\bar{m}_1^2)\right)-n_F\left(p_0+E_k(\bar{m}_1^2)\right) \sim p_0 n_F'\left(E_k(\bar{m}_1^2)\right)$. Thus, one is led to
\begin{align}
  \mathcal{F}_1'(q_-,k,|\bm p|;\bar{m}_f^2,\bar{m}_f^2)=&(2k-|\bm p|)\frac{n_F'\left(E_k(\bar{m}_f^2)\right)}{8E_k(\bar{m}_f^2)}p_0+\mathcal{O}(p_0^2)\,.  \label{eq:F1pri3}
\end{align}
Combining \Eq{eq:F1pri2} and the relation as follows
\begin{align}
    &\mathrm{Im} I_{2,k}(p\,;\bar{m}_1,\bar{m}_2;\mathcal{FF}_{(2,1)})=-\frac{\partial}{\partial \bar{m}_1^2}\mathrm{Im} I_{2,k}(p\,;\bar{m}_1,\bar{m}_2;\mathcal{FF}_{(1,1)})\,,
\end{align}
one is able to obtain 
\begin{align}
    &\mathrm{Im} I_{2,k}(p\,;\bar{m}_f^2,\bar{m}_f^2;\mathcal{FF}_{(2,1)})\nonumber\\[2ex]
=&\Bigg[\frac{(k-|\bm p|)}{k|\bm p|}\frac{n_F'\left(E_k(\bar{m}_f^2)\right)}{8E_k(\bar{m}_f^2)}+(2k-|\bm p|)\frac{n_F'\left(E_k(\bar{m}_f^2)\right)}{16E_k(\bar{m}_f^2)^3}-(2k-|\bm p|)\frac{n_F''\left(E_k(\bar{m}_f^2)\right)}{16E_k(\bar{m}_f^2)^2}\Bigg]p_0+\mathcal{O}(p_0^2)\,, \label{eq:I2-FF21}
\end{align}
where $\bar{m}_1=\bar{m}_2=\bar{m}_f$ has been used. From \Eq{eq:I2-FF21} one observes that 
\begin{align}
 \mathrm{Im} I_{2,k}(p\,;\bar{m}_f^2,\bar{m}_f^2;\mathcal{FF}_{(2,1)})\sim \frac{1}{|\bm p|} \,,  \label{}
\end{align}
in the limit of $|\bm p| \to 0$.

\end{document}